\documentclass[10pt,journal]{IEEEtran}
\usepackage{amsmath,amsfonts}
\usepackage{array}
\usepackage[caption=false,font=normalsize,labelfont=sf,textfont=sf]{subfig}
\usepackage{textcomp}
\usepackage{stfloats}
\usepackage{url}
\usepackage{verbatim}
\usepackage{float}
\usepackage{graphicx}
\usepackage{algorithm}
\usepackage{algpseudocode}
\def\BibTeX{{\rm B\kern-.05em{\sc i\kern-.025em b}\kern-.08em
    T\kern-.1667em\lower.7ex\hbox{E}\kern-.125emX}}
\usepackage{balance}
\begin{document}
\title{RepuNet: A Reputation System for Mitigating Malicious Clients in DFL}
\author{Isaac Marroqui Penalva, Enrique Tomás Martínez Beltrán, Manuel Gil Pérez, Alberto Huertas Celdrán
\thanks{Isaac Marroqui Penalva, Enrique Tomás Martínez Beltrán, Manuel Gil Perez, and Alberto Huertas Celdrán are with the University of Murcia, Spain.}}

\maketitle

\begin{abstract}
Decentralized Federated Learning (DFL) enables nodes to collaboratively train models without a central server, introducing new vulnerabilities since each node independently selects peers for model aggregation. Malicious nodes may exploit this autonomy by sending corrupted models (model poisoning), delaying model submissions (delay attack), or flooding the network with excessive messages, negatively affecting system performance. Existing solutions often depend on rigid configurations or additional infrastructures such as blockchain, leading to computational overhead, scalability issues, or limited adaptability. To overcome these limitations, this paper proposes RepuNet, a decentralized reputation system that categorizes threats in DFL and dynamically evaluates node behavior using metrics like model similarity, parameter changes, message latency, and communication volume. Nodes' influence in model aggregation is adjusted based on their reputation scores. RepuNet was integrated into the Nebula DFL platform and experimentally evaluated with MNIST and CIFAR-10 datasets under non-IID distributions, using federations of up to 25 nodes in both fully connected and random topologies. Different attack intensities, frequencies, and activation intervals were tested. Results demonstrated that RepuNet effectively detects and mitigates malicious behavior, achieving F1 scores above 95\% for MNIST scenarios and approximately 76\% for CIFAR-10 cases. These outcomes highlight RepuNet’s adaptability, robustness, and practical potential for mitigating threats in decentralized federated learning environments.
\end{abstract}

\begin{IEEEkeywords}
Federated Learning, reputation systems, malicious node detection, anomaly detection, reputation-based aggregation, attack mitigation, adversarial behavior in DFL
\end{IEEEkeywords}

\section{Introduction}

\IEEEPARstart{F}{ederated} Learning (FL)~\cite{mcmahan2017communication} has emerged as a key solution in the era of Artificial Intelligence (AI) to train models collaboratively without compromising data privacy. Traditionally, centralized FL systems have dominated this field, but their reliance on a single server introduces security vulnerabilities, as this central point can become a target for malicious attacks. To mitigate these risks, DFL has gained relevance \cite{10251949}, eliminating the need for a central server, removing the single point of failure, and enhancing system robustness. This is crucial in sensitive domains such as healthcare or finance.
However, DFL introduces unique security challenges due to the lack of central supervision during training. In the first phase, nodes train local models independently, which makes them vulnerable to \textit{model poisoning} attacks, where adversaries inject distorted updates to degrade global convergence~\cite{9878267}. In the second phase, nodes exchange parameters with neighbors, a stage exposed to \textit{communication delay} attacks. These can cause desynchronization and unstable learning dynamics, as outdated models propagate through the network~\cite{10304380}. Finally, during aggregation, malicious participants can launch \textit{flooding or DoS} attacks by overwhelming the network with excessive messages, disrupting communication and slowing convergence~\cite{9454328}. These threats call for robust mechanisms to mitigate malicious behavior.
There are various approaches to mitigating these attacks in DFL, such as the use of blockchain \cite{10480262}, cluster-based aggregations \cite{PANIGRAHI2023108900}, suspicious model filtering \cite{gabrielli2024protectingfederatedlearningextreme}, and cryptographic defense techniques \cite{EURECOM+6974}. However, these approaches have certain limitations. The effectiveness of cluster-based systems depends on proper configuration, making them vulnerable to manipulation if a group contains a high proportion of attackers. Furthermore, nodes may coordinate to influence reputation within their own groups, avoiding detection. If hierarchical aggregation is later performed, additional computational overhead is introduced, which may not be well-suited for decentralized environments. In the case of blockchain usage, while it provides traceability and resistance to manipulation, it introduces high computational and storage costs, along with potential scalability issues as the number of nodes in the network grows. Model filtering can be too restrictive or generate false positives, affecting learning diversity. On the other hand, cryptographic solutions, although effective in some contexts, add computational overhead, making them unfeasible in resource-limited scenarios.
To address these limitations, this paper proposes RepuNet, a decentralized reputation system specifically designed for DFL. RepuNet serves as a dynamic mechanism for detecting and mitigating attacks by continuously evaluating the behavior of nodes over time. It penalizes those that exhibit anomalous patterns—such as communication delays, transmission of inconsistent parameters, or excessive traffic generation. By integrating reputation scores into each training round, the system can exclude malicious node contributions without significantly affecting network stability, thereby enhancing both security and global model convergence.

\noindent \textbf{The main contributions of this work are as follows:}
\begin{itemize}
    \item \textbf{Design of a threat-aware and adaptive reputation system for DFL.} This work defines a threat model that identifies and categorizes key attacks in DFL—such as model poisoning, intentional communication delays, and flooding—and uses it as the basis for RepuNet, a decentralized reputation mechanism. The system evaluates the behavior of neighboring nodes using dynamic metrics (e.g., model similarity, parameter change ratio, latency, and message volume), enabling adaptive weighting and detection of malicious patterns.

    \item \textbf{Progressive penalization and controlled exclusion mechanism.} RepuNet gradually reduces the influence of nodes with low reputation in the aggregation process without disconnecting them. This supports the detection of persistent adversarial behavior while allowing reintegration if behavior improves in later rounds.

    \item \textbf{Deployment of RepuNet on the Nebula platform under realistic DFL conditions.} The system has been implemented and evaluated using the Nebula platform~\cite{nebula2025}, which supports configurable topologies, attack scenarios, and real-time metric visualization. Experiments were performed in emulated environments with up to 25 nodes, fully connected and random topologies, and three types of attacks. The evaluation used MNIST and CIFAR-10 datasets under non-IID distributions, demonstrating significant improvements in anomaly detection, robust behavior under different levels of adversarial activity, and stable model performance across rounds.
\end{itemize}

The structure of this article is as follows: Section~\ref{sec:related} reviews related work on attack mitigation in DFL. Section~\ref{sec:threatmodel} introduces the threat model and explains the rationale for focusing on specific attacks. Section~\ref{sec:RepuNet} details the design and operation of the proposed reputation system, RepuNet. Section~\ref{sec:validation} describes the experimental setup and attack scenarios. Section~\ref{sec:discussion} presents and analyzes the results. Finally, Section~\ref{sec:conclusion} summarizes the main contributions and outlines future work.

\section{Related Work}
\label{sec:related}

This section reviews reputation systems in FL and defense mechanisms against malicious behavior in decentralized networks. It covers blockchain-based FL, hierarchical aggregation, and attack detection strategies—such as data and model poisoning or DDoS—as well as both centralized and decentralized approaches that promote honest participation and penalize adversarial nodes.

\subsection{Reputation Systems in Federated Learning}

A centralized approach \cite{9428537} updates reputation scores using behavioral history and a beta distribution. Nodes must exceed an initial reputation threshold to participate, ensuring reliability from the outset.
Decentralized systems offer transparency and resilience. \cite{8832210} stores contributions on blockchain; reputation uses direct and indirect feedback. \cite{zhao2019mobile} applies blockchain reputation in IoT-FL to promote cooperation.
Cooperative incentives have emerged. \cite{domingo2021secure} proposes a co-utility rule system to reward compliant behavior without central coordination. \cite{panigrahi2023reputation} explores hierarchical aggregation using accuracy and participation to select models.
\cite{gao2022fgfl} proposes FGFL, a blockchain system using reputation and contribution metrics to adjust incentives.
Together, these approaches reflect a shift from static, centralized schemes toward adaptive, incentive-aware, and decentralized reputation frameworks suitable for heterogeneous federated learning environments.

\subsection{Attack Mitigation Strategies}

DFL is vulnerable to various attacks that compromise the integrity, efficiency, or convergence of the learning process. These include model manipulation, flooding, and message delays.
Several studies address model poisoning and data manipulation. \cite{inproceedings} studies poisoning in intrusion detection and proposes model filtering. In smart grid contexts, \cite{9878267} employs deep learning to detect falsified data without compromising privacy.
To mitigate network saturation, \cite{9454328} proposes a decentralized FL-based architecture for anomaly detection in industrial IoT. Likewise, \cite{10186438} demonstrates how distributed traffic analysis in IoT-Edge environments can identify malicious patterns while preserving data locality.
Delay attacks—where nodes transmit outdated updates to desynchronize the network—are addressed in \cite{10304380}, which uses hierarchical FL to reduce the influence of slower participants. \cite{dfl-security-privacy} surveys anomaly detection, cryptography, and reputation mechanisms.
Blockchain has also been proposed to enhance trust and traceability. For example, \cite{article} uses sharding and layered validation to secure aggregation, while \cite{10.1145/3703631} targets passive threats like “lazybone” nodes through multidimensional evaluation of contribution quality.
These strategies highlight the need for hybrid defenses combining anomaly detection, contribution filtering, decentralized coordination, and adaptive reputation, to ensure resilience in diverse DFL scenarios.

\subsection{Comparison with Existing Approaches}

RepuNet allows decentralized reputation where nodes assess neighbors without central authority or blockchain. Reputation evolves dynamically from local metrics such as model similarity, parameter variation, latency, and message volume, using a neutral initial value to avoid bias.
Unlike static or irreversible schemes, RepuNet allows nodes to recover reputation through consistent behavior, balancing penalization and reintegration.
Compared to blockchain-based systems, RepuNet avoids the heavy computational and communication overhead. It eliminates the need for immutable global records and relies on lightweight, adaptive updates performed locally at each node. Additionally, peer feedback can be optionally incorporated to strengthen individual assessments, especially in scenarios with partial observations, newly joined nodes, or sparse topologies. This combination enhances system resilience without compromising decentralization.
As shown in Table~\ref{tab:comparison-unified}, RepuNet integrates more metrics—including communication anomalies—and avoids overhead from blockchain-based systems.

\begin{table*}[htbp]
\centering
\caption{Comparison of anomaly detection criteria, detected attacks, and mitigation strategies in FL systems.}
\renewcommand{\arraystretch}{1.2}
\resizebox{\textwidth}{!}{%
\begin{tabular}{
    >{\raggedright\arraybackslash}m{2cm}   
    >{\centering\arraybackslash}m{2.5cm}       
    >{\centering\arraybackslash}m{7cm}     
    >{\centering\arraybackslash}m{1cm}     
    >{\centering\arraybackslash}m{2.5cm}     
    >{\centering\arraybackslash}m{2cm}     
    >{\centering\arraybackslash}m{6.8cm}     
}
\hline
\textbf{Work} (Year) & \textbf{FL Type} & \textbf{Detection / Penalization Criteria} & \textbf{Feedback} 
& \textbf{Detected Attack(s)} & \textbf{Anomaly Detection} & \textbf{Action Taken} \\
\hline
\cite{zhao2019mobile} (2019) & DFL (Blockchain) & Crowdsourced contribution history on blockchain & X 
& -- & X & Trustworthy nodes rewarded; reputation affects participation \\

\cite{8832210} (2019) & DFL (Blockchain) & Trust via contribution quality and peer reputation (direct/indirect) & \checkmark 
& -- & X & Adjusts reputation; low scores discourage future inclusion \\

\cite{inproceedings} (2019) & IoT & Model deviation from expected behavior (filtering) & X 
& Poisoning & \checkmark & Suspicious updates discarded before aggregation \\

\cite{domingo2021secure} (2021) & DFL (Blockchain) & Violation of cooperation rules (co-utility model) & X 
& -- & X & Rewards or penalties applied via rule compliance \\

\cite{9428537} (2022) & CFL & Deviation in behavioral history (beta-based reputation) & X 
& -- & X & Reputation score adjusted (centralized); no exclusion \\

\cite{gao2022fgfl} (2022) & DFL (Blockchain) & Quality and frequency of valid contributions & X 
& Lazy / low-effort participation & \checkmark & Contributors receive dynamic rewards or are excluded if passive \\

\cite{9878267} (2022) & Smart Grid & Anomalous data pattern detection using deep learning & X 
& FDIA (false data injection) & \checkmark & Classifier-based filtering of malicious updates \\

\cite{9454328} (2022) & Industrial IoT & Traffic anomalies detected via federated classifiers & X 
& DDoS & \checkmark & Detection used to alert and reconfigure network paths \\

\cite{10186438} (2023) & IoT Edge & Distributed traffic anomaly pattern detection in IoT & X 
& DDoS, poisoning & \checkmark & Local anomaly response; contributes to distributed decision-making \\

\cite{panigrahi2023reputation} (2023) & Hierarchical CFL & Accuracy, consistency, and past participation history & X 
& -- & X & Contribution filtered before aggregation based on reputation \\

\cite{dfl-security-privacy} (2024) & DFL (Blockchain) & Overview of multiple metrics: gradients, consistency, etc. (survey) & X 
& Poisoning, backdoor, Sybil, free-riding & \checkmark & Describes possible responses: filtering, exclusion, weight adjustment \\

\cite{10304380} (2024) & FL (Hierarchical) & Model submission latency (ranking-based classification) & X 
& Delay & X & Low-ranked nodes deprioritized or excluded in hierarchy \\

\cite{article} (2025) & DFL (Blockchain) & Distributed model validation through consensus (sharded blockchain) & X 
& Sybil, poisoning & \checkmark & Invalid models rejected during consensus verification \\

\cite{10.1145/3703631} (2025) & CFL & Contribution scoring via loss, gradient norm, and activity level & \checkmark 
& Lazybone & \checkmark & Low-contributing nodes excluded from aggregation \\

\textbf{RepuNet} (2025) & DFL & Similarity, parameter change, arrival latency, message volume & \checkmark 
& Poisoning, delay, flooding & \checkmark & Dynamic penalization of reputation, exclusion from aggregation, and adaptive reintegration \\
\hline
\end{tabular}
}
\label{tab:comparison-unified}
\end{table*}

\section{Threat Model}
\label{sec:threatmodel}

DFL faces several vulnerabilities due to the autonomy of participating nodes, which may act maliciously and compromise the quality, efficiency, or stability of the collaborative process.

The most relevant threats in DFL can be grouped into four categories:

\begin{itemize}
\item \textbf{Model-based attacks:} manipulation of local parameters (e.g., \textit{model poisoning}, \textit{backdoor attacks}).
\item \textbf{Communication-based attacks:} such as \textit{delayer attacks}, \textit{flooding}, or DoS.
\item \textbf{Data-based attacks:} noise injection, false data submission, or inference of private information.
\item \textbf{Identity/participation attacks:} including \textit{Sybil}, \textit{free-riding}, or \textit{lazybone} behaviors.
\end{itemize}

This work targets three threats due to their impact and feasibility of local detection:

\begin{itemize}
\item \textbf{Model poisoning:} injection of malicious updates to degrade the global model.
\item \textbf{Delayer attack:} intentional delay in sending models to disrupt synchronization.
\item \textbf{Flooding attack:} excessive message generation to saturate the network.
\end{itemize}

These attacks degrade model quality, synchronization, or communication, and can be detected using local metrics such as model similarity, latency, or message count. They are common in real-world deployments.
Other threats (e.g., inference, Sybil, or passive participation) require identity validation or secure exchanges and are beyond the scope of this work. By focusing on locally observable threats, the proposed system remains practical and suitable for controlled experimental validation, as described in Section~\ref{sec:validation}.

\section{RepuNet: An Adaptive Reputation System to Mitigate Attacks in DFL}
\label{sec:RepuNet}

This section presents RepuNet, a decentralized and adaptive reputation system specifically designed to detect and mitigate attacks in DFL. RepuNet continuously evaluates the behavior of participating nodes using a set of observable metrics and adjusts their influence in model aggregation according to their reputation. This section details the reputation scoring mechanism, the metrics used, the step-by-step operational flow, the reputation update process, exclusion and recovery criteria, and its practical integration in the Nebula experimental platform.

\subsection{Reputation System Foundations}

RepuNet is designed for fully DFL environments with the goal of evaluating node behavior to detect and mitigate malicious activity while preserving contribution quality and minimizing overhead.
Each node autonomously decides to accept, weight, or exclude incoming models. Its design emphasizes training efficiency and system stability, avoiding bottlenecks or mechanisms that could hinder global convergence. This enables a robust and resilient collaborative environment, even in the presence of anomalous behavior.

\subsection{Decentralized Architecture and Operation}

\begin{figure}[htbp]
    \centering
    \includegraphics[width=0.5\textwidth]{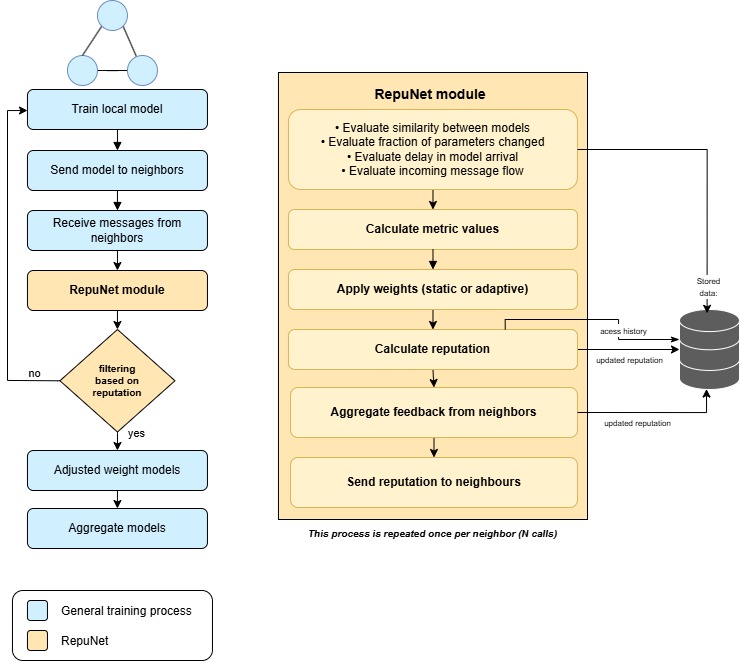}
    \caption{Interaction between a federated node and the RepuNet module, showing the flow of training, reputation evaluation, and aggregation.}
    \label{fig:reputation_flow}
\end{figure}

RepuNet is modular and decentralized. Each node in the federated network includes an autonomous reputation component that evaluates the behavior of its neighbors during each training round. This evaluation determines whether received models should be filtered or adjusted before aggregation, depending on the sender’s reputation.
Figure~\ref{fig:reputation_flow} shows two main components. On the left is the standard local training flow: local model training, model broadcasting, reception of external models, filtering based on reputation, and aggregation. On the right, the internal RepuNet module calculates a reputation score for each neighbor using observable behavioral metrics. The outcome of this evaluation guides the filtering and adjustment decisions.
Upon receiving external models, the node calls the RepuNet module. This module evaluates each neighbor based on metrics such as model similarity, fraction of parameters changed, delay in model arrival, and incoming message flow. These metrics are processed and weighted—either statically or adaptively—to produce a reputation score. Feedback from other nodes is also aggregated to enhance the evaluation. Reputations are stored locally and then disseminated to neighbors in each round as part of the protocol.
The updated reputation influences the filtering decision: if the reputation is below a given threshold, the model is discarded; otherwise, it is incorporated with a weight proportional to its reputation. The evaluation is performed once per neighbor per round.

The complete operational flow is summarized below:

\begin{enumerate}
    \item \textbf{Local training:} The node updates its model using local private data.
    \item \textbf{Model broadcast:} The model is shared with neighboring nodes.
    \item \textbf{Message reception:} Models are received from neighbors.
    \item \textbf{Reputation evaluation:} RepuNet evaluates each neighbor using local metrics and aggregated feedback.
    \item \textbf{Filtering decision:} If a neighbor’s reputation is too low, the model is discarded.
    \item \textbf{Adjusted weighting:} Accepted models are weighted based on the sender's reputation.
    \item \textbf{Aggregation:} Reputation-weighted models are aggregated with the local model.
    \item \textbf{Reputation dissemination:} Updated reputation scores are stored and sent to all neighbors.
\end{enumerate}

Unlike binary filters, RepuNet adjusts each model’s aggregation weight. This adaptive integration allows the contribution of each node to evolve dynamically based on behavioral history. The next sections detail the computation of reputation and the metrics involved. The operational logic can be formalized through Algorithm~\ref{alg:fedrep}, which summarizes the reputation-guided federated training cycle.

\begin{algorithm}
\scriptsize
\caption{\scriptsize Federated round with reputation-guided adaptive aggregation}
\label{alg:fedrep}
\begin{algorithmic}[1]
\State \textbf{COMMUNICATION\_ROUND()}
\State \hspace{0.5cm} \textbf{TrainLocalModel()}
\State \hspace{0.5cm} \textbf{SendModelToNeighbors()}
\State \hspace{0.5cm} $messages \gets$ \textbf{ReceiveMessagesFromNeighbors()}
\For{each $message$ in $messages$}
    \State $rep \gets$ \textbf{GetReputation}($message.sender$)
    \If{\textbf{IsReputationSufficient}($rep$)}
        \State $weight \gets$ \textbf{ComputeWeightFromReputation}($rep$)
        \State \textbf{StoreAcceptedModel}($message.model$, $weight$)
    \EndIf
\EndFor
\State \textbf{AggregateAcceptedModels()}
\State
\State \textbf{UPDATE\_REPUTATION()}
\For{each $neighbor$}
    \State $metrics \gets$ \textbf{CalculateMetrics}($neighbor$)
    \State $score \gets \sum_j weights[j] \cdot NormalizeMetric(metrics[j])$
    \State $history \gets$ \textbf{GetReputationHistory}($neighbor$)
    \State $rep \gets \sum_{r \in history} \omega^{(r)} \cdot rep[r] + \omega^{(t)} \cdot score$
    \If{feedback available}
        \State $avg\_fb \gets$ average reputation received from neighbors
        \State $rep \gets \eta \cdot rep + (1 - \eta) \cdot avg\_fb$
    \EndIf
    \State \textbf{StoreReputation}($neighbor$, $rep$)
    \State \textbf{SendReputationToNeighbor}($neighbor$, $rep$)
\EndFor
\end{algorithmic}
\end{algorithm}

\subsection{Metrics Used in RepuNet}

RepuNet relies on four primary metrics to evaluate the behavior of neighbors in a DFL network. These metrics are divided into two categories: \emph{model quality} and \emph{communication behavior}. Each metric returns a normalized value between 0 and 1, where 1 indicates optimal behavior. The system applies soft penalties to deviations, and the metrics are combined with dynamic weights to generate the reputation score.

\subsubsection{Model Similarity}

This metric assesses the alignment between the local model and the one received. Four classic distance measures are used: cosine, Euclidean, Manhattan, and Pearson correlation. The final similarity is a weighted average:
{\small \begin{align}
\text{similarity} = \sum_k \gamma_k \cdot S_k
\end{align} }
where \( S_k \) represents each individual measure and \( \gamma_k \) its weight. Low values indicate anomalous deviations in model parameters.

\subsubsection{Fraction of Parameters Changed}

RepuNet computes this metric by maintaining a history of change fraction values ($f$) and their associated thresholds ($t$), from which the mean and standard deviation are derived:
{\small \begin{align}
\mu_f &= \frac{1}{N} \sum_{i=0}^{N} f_i, &
\sigma_f &= \sqrt{\frac{1}{N} \sum_{i=0}^{N} (f_i - \mu_f)^2} \\
\mu_t &= \frac{1}{N} \sum_{i=0}^{N} t_i, &
\sigma_t &= \sqrt{\frac{1}{N} \sum_{i=0}^{N} (t_i - \mu_t)^2}
\end{align} }
From these statistics, acceptable upper limits are defined:
{\small \begin{align}
\text{limit}_f = (\mu_f + \sigma_f) \times 1.05, \quad
\text{limit}_t = (\mu_t + \sigma_t) \times 1.10
\end{align} }
If the current value exceeds the corresponding threshold, it is considered anomalous. A penalty proportional to the deviation is then calculated:
{\small \begin{align}
P = \frac{|f_{\text{current}} - \mu_f|}{\mu_f}
\end{align} }
This penalty is transformed into a smoothed score using a sigmoid function:
{\small \begin{align}
S = 1 - \left( \frac{1}{1 + e^{-P}} \right)
\end{align} }
If no anomaly is detected, the score is perfect: $S = 1.0$. The scores for the change fraction ($S_f$) and threshold ($S_t$) are combined with equal weights:
{\small \begin{align}
f_{\text{final}} = 0.5 \cdot S_f + 0.5 \cdot S_t
\end{align} }
To avoid overpenalizing isolated variations, a temporal smoothing mechanism is applied:
{\small \begin{align}
f^{(t)}_{\text{final}} = \lambda^{(t)} \cdot f^{(t)}_{\text{actual}} + (1 - \lambda^{(t)}) \cdot f^{(t-1)}_{\text{final}}
\end{align} }
where $ \lambda^{(t)} \in [0, 1] $ is a sensitivity coefficient for recent changes.

Finally, if a node does not receive a model from a neighbor during an iteration, this metric is automatically penalized by reducing the score by 50\% from its previous value.

\subsubsection{Model Arrival Latency}

This metric evaluates how promptly a neighboring node delivers its model in each communication round. Its goal is to detect \textbf{anomalous delays} that may affect node synchronization and aggregation stability.

The computation begins by logging the model arrival time \( t_{\text{arrival}} \), compared to the start time of the round. Depending on the temporal origin of the model, latency is defined as:
{\small \begin{align}
\text{latency} = 
\left\{
\begin{array}{ll}
t_{\text{arrival}} - t_{\text{start\_current\_round}} & \text{current round model} \\[0.6ex]
t_{\text{arrival}} - t_{\text{start\_model\_round}} & \text{previous round model}
\end{array}
\right.
\end{align} }
Based on past observations, the historical latency average is computed as:
{\small \begin{align}
\overline{\text{latency}} = \frac{1}{N} \sum_{i=1}^{N} \text{latency}_i
\end{align} }
When the attack begins from the first iteration, round 0 is used as the baseline, and the tolerance threshold is set at 150\% of the average latency. The deviation is calculated as the difference between the current latency and the historical average latency.
The final score is assigned as follows:
{\small \begin{align}
\text{score} =
\left\{
\begin{array}{ll}
1.0 & \text{if within threshold} \\[1.2ex]
\displaystyle \frac{1}{1 + \exp\left( \frac{|\Delta t|}{\tau^{(t)}} \right)} & \text{if above}
\end{array}
\right.
\end{align} }
where \( \Delta t \) is the observed deviation from the historical mean and \( \tau^{(t)} \) is a tunable temporal tolerance parameter.
To smooth inter-round variation, an exponential update is applied:
{\small \begin{align}
\bar{s}^{(t)}_{\text{lat}} = \mu^{(t)} \cdot s^{(t)}_{\text{lat}} + (1 - \mu^{(t)}) \cdot \bar{s}^{(t-1)}_{\text{lat}}
\end{align} }
In early rounds, where history is limited, a bootstrapping penalty is introduced:
{\small \begin{align}
\bar{s}^{(t)}_{\text{lat}} = s^{(t)}_{\text{lat}} \cdot (1 - \delta^{(t)}), \quad \delta^{(t)} \in [0, 1], \text{ typically } \delta^{(t)} = 0.05
\end{align} }
If no model is received from a node during a round, its score is reduced by 50\,\% from the previous round.

This metric helps detect both intentional delays and \textit{lack of participation}, contributing to the preservation of temporal coherence in the federation.

\subsubsection{Incoming Message Flow}

This metric evaluates the communicative activity of a node by comparing the number of messages sent to a specific neighbor during the current round with the collective behavior of the previous round. Its goal is to detect anomalies like excessive or persistent overcommunication typical of flooding attacks.

For each node pair $(a, b)$, the number of messages sent in round $r$ is denoted $m_r^{(a,b)}$, and the previous round $(r-1)$ message counts across all pairs $(i, j)$ are collected. From this, the 25th percentile $P_{25}$ is computed as a reference, along with standard deviation $\sigma_{r-1}$ and mean $\mu_{r-1}$.

The relative increase is defined as:
{\small \begin{align}
\text{rel\_incr} = \max\left( \frac{m_r^{(a,b)} - P_{25}}{P_{25}}, \; 0 \right)
\end{align} }
A dynamic tolerance margin is calculated based on variability:
{\small \begin{align}
\text{dynamic\_margin} = \frac{\sigma_{r-1} + 1}{\log(1 + P_{25}) + 1}
\end{align} }

If the relative increase exceeds the margin, a smoothed logarithmic exponential penalty is applied:
{\small \begin{align}
s^{(r)}_{\text{msg}} \leftarrow s^{(r)}_{\text{msg}} \cdot 
\exp\left( 
    -\left( 
        \frac{\log\left(1 + \text{rel\_incr} - \text{dynamic\_margin}\right)}{
            \log\left(1 + \text{dynamic\_margin}\right) + \varepsilon
        }
    \right)^2
\right)
\end{align}}

If the node also exceeds the global mean:
{\small \begin{align}
\text{amplification} = 1 + \frac{increase\_mean}{\mu_{r-1} + \varepsilon}
\end{align} }
{\small \begin{align}
s^{(r)}_{\text{msg}} \leftarrow s^{(r)}_{\text{msg}} \cdot 
\exp\left( -(\text{extra\_penalty} \cdot \text{amplification})^2 \right)
\end{align} }
where increase\_mean adjusts automatically based on the system phase.

A per-node and per-neighbor temporal history is maintained. Repeated low scores across rounds trigger additional multiplicative penalties.
To avoid permanent exclusion, a minimum adaptive bound is enforced:
{\small \begin{align}
s^{(r)}_{\text{msg}} \leftarrow \max\left( s^{(r)}_{\text{msg}}, f_{\text{min}}(r, a, b) \right)
\end{align} }

Values are smoothed with a weighted average of up to three prior rounds.
This formulation detects and penalizes meaningful communication deviations proportionally and dynamically, without requiring fixed thresholds.

These four metrics form the foundation of the RepuNet reputation system and are summarized in Table~\ref{tab:reputation_metrics}.

\begin{table*}[htbp]
\centering
\caption{Summary of metrics used in reputation computation}
\renewcommand{\arraystretch}{1.1}
\label{tab:reputation_metrics}
\resizebox{\textwidth}{!}{%
\begin{tabular}{
    >{\raggedright\arraybackslash}m{4cm}
    >{\centering\arraybackslash}m{1.3cm}
    >{\centering\arraybackslash}m{4.2cm}
    >{\centering\arraybackslash}m{1cm}
    >{\centering\arraybackslash}m{1cm}
    >{\centering\arraybackslash}m{3.2cm}
    >{\centering\arraybackslash}m{6.8cm}
}
\hline
\textbf{Metric} & \textbf{Type} & \textbf{Purpose} & \textbf{Symbol} & \textbf{Range} & \textbf{Penalty trigger} & \textbf{Normalization} \\
\hline
Model similarity & Model quality & Align local and received models & $S$ & $[0, 1]$ & Low similarity & Weighted average (cosine, Euclidean, Manhattan, Pearson) \\
\hline
Fraction of parameters changed & Model quality & Detect abrupt model changes & $F$ & $[0, 1]$ & High deviation from past & Sigmoid penalty on deviation \\
\hline
Model arrival latency & Communication & Penalize delayed models & $L$ & $[0, 1]$ & Above historical latency & Sigmoid based on delay deviation \\
\hline
Incoming message flow & Communication & Detect flooding behavior & $M$ & $[0, 1]$ & High message volume & Exponential penalty vs. dynamic margin \\
\hline
\end{tabular}
}
\end{table*}

\subsection{Weight Assignment and Reputation Computation}

Once the metrics for each neighbor have been computed and normalized, RepuNet assigns dynamic weights that reflect the discriminative power of each metric in the current round. This assignment is performed individually per node, based on deviation from a reference historical behavior.
If the system has a stable enough history, deviations are computed with respect to the accumulated historical mean. In attack scenarios activated in later rounds (e.g., after round 7), the prior history is used as the reference pattern. In contrast, if the attack begins in the first iteration, only the data from round 0 is used as a baseline.
Let \( m_j^{(t)} \) be the observed value of metric \( j \) for a node in round \( t \), and \( \mu_j \) its historical mean. The absolute deviation is calculated as:
{\small \begin{align}
d_j^{(t)} = \left| m_j^{(t)} - \mu_j \right|
\end{align} }
These deviations are aggregated into a total value:
{\small \begin{align}
D^{(t)} = \sum_k d_k^{(t)}
\end{align} }
and used to compute the normalized weight assigned to each metric:
{\small \begin{align}
w_j^{(t)} = \frac{d_j^{(t)}}{D^{(t)}}
\end{align} }
If all deviations are zero (i.e., the node behaves exactly as expected), randomly normalized weights are assigned to prevent stagnation in the reputation process. Additionally, a dynamic lower bound is applied to ensure no relevant metric is ignored.
Once the weights are set, an intermediate reputation score \( S_i^{(t)} \) is computed for each neighbor \( i \), using a weighted sum of the metrics:
{\small \begin{align}
S_i^{(t)} = \sum_j w_j^{(t)} \cdot m_j^{(t)}
\end{align} }
This score is used to update the node’s overall reputation, taking into account the history of past rounds \( \mathcal{H}_i^{(t)} \). The reputation is updated as a weighted combination of past values and the current score:
{\small \begin{align}
R_i^{(t)} = \sum_{r \in \mathcal{H}^{(t)}_i} \omega^{(r)} \cdot R_i^{(r)} + \omega^{(t)} \cdot S_i^{(t)}
\end{align} }
where \( \omega^{(r)} \) and \( \omega^{(t)} \) are weights satisfying \( \sum \omega = 1 \), allowing balance between historical behavior and current observation.
Finally, if reputation feedback is enabled, RepuNet can incorporate the opinions of neighboring nodes. Given a set of received values \( R_{j \rightarrow i}^{(t)} \), an adjusted reputation is computed using a convex combination controlled by a parameter \( \eta \in [0,1] \):
{\small \begin{align}
\tilde{R}_i^{(t)} = \eta \cdot R_i^{(t)} + (1 - \eta) \cdot \text{mean}\left( \left\{ R_{j \rightarrow i}^{(t)} \right\} \right)
\end{align} }
The resulting value \( \tilde{R}_i^{(t)} \) represents the final reputation of the node for that round and will be used to determine its influence in the next aggregation iteration.

\subsection{Aggregation, Exclusion, and Recovery Control}

Once the final reputation of each node has been computed, RepuNet uses this value to make local decisions about including or excluding their models in the aggregation process. If a neighbor's reputation falls below a predefined trust threshold, their model is automatically discarded for that round. This exclusion prevents anomalous, malicious, or inconsistent behaviors from influencing the global model.
It is important to note that exclusion does not imply disconnection or blocking of the node in the system: the penalty only applies to the use of its model, allowing continued participation in subsequent rounds and re-evaluation. This ensures the architecture remains open and tolerant toward nodes that may recover their behavior after temporary penalties.
The exclusion threshold is configurable and can be adapted to the nature of the environment or the attack. In typical configurations, reputation values below 0.6 lead to model rejection, while higher reputations allow the node’s contribution to be weighted accordingly.
Furthermore, reputation is evaluated and updated in every round. If a node improves its behavior and its metrics begin to align with the expected pattern, its reputation gradually increases, allowing for natural and progressive reintegration into aggregation. This recovery property avoids permanent penalties for transient issues or isolated failures.
Overall, this control mechanism provides a flexible, adaptive, and resilient system that acts as a dynamic filter against untrustworthy or malicious contributions while encouraging the reintegration of rehabilitated nodes.

\subsection{Distributed Feedback and Robustness}

In addition to local metrics, RepuNet allows each node to incorporate reputation feedback from trusted neighbors. This complementary information reinforces the individual assessment, especially in scenarios where a node may lack sufficient observations or its perception may be distorted due to specific network conditions or atypical behavior.
Each node may receive reputation scores issued by its neighbors in the current round. This feedback is incorporated through a weighted combination of the local evaluation \( R_i^{(t)} \) and the average of the received evaluations \( \text{mean}(\{ R^{(t)}_{j \rightarrow i} \}) \), regulated by a coefficient \( \eta \in [0,1] \) that controls the level of trust in the internal assessment:
{\small \begin{align}
\tilde{R}_i^{(t)} = \eta \cdot R_i^{(t)} + (1 - \eta) \cdot \text{mean}\left( \{ R^{(t)}_{j \rightarrow i} \} \right)
\end{align}}
Since only nodes with an active reputation module are allowed to issue evaluations, it is guaranteed that all received scores originate from valid and trustworthy participants, thereby preventing the propagation of manipulated information.
This feedback mechanism serves as a decentralized consensus approach, enabling the smoothing of isolated evaluation errors, mitigating local fluctuations, and reinforcing system robustness against targeted attacks or biased perceptions. This preserves the stability and adaptability of the federated process.
In subsequent rounds, the system repeats the same evaluation cycle: metric computation, dynamic weight assignment, reputation update, neighbor feedback incorporation, and dissemination of the final reputation. This continuous and adaptive mechanism allows the federated system to remain resilient against malicious nodes while supporting the reintegration of nodes that recover reliable behavior.

\subsection{Integration of RepuNet into the Nebula Platform}

RepuNet has been implemented as a modular, autonomous, and platform-independent component. For experimental evaluation, it was integrated into the \textit{Nebula} platform, which supports decentralized FL scenarios and includes real-time visualization via TensorBoard~\cite{nebula2025}. The integration enables testing across topologies and attack types.

Nebula’s architecture consists of three main layers:

\begin{itemize}
    \item \textbf{Frontend:} a graphical interface where users can:
    \begin{itemize}
        \item Activate or deactivate RepuNet.
        \item Select evaluation metrics and their weighting scheme.
        \item Configure parameters such as the exclusion threshold, feedback weight \(\eta\), and temporal smoothing.
    \end{itemize}

    \item \textbf{Controller:} an intermediate layer that propagates configurations to all nodes, ensuring consistency across rounds.

    \item \textbf{Core:} the execution layer for each node, which handles:
    \begin{itemize}
        \item Local training on private data.
        \item Model exchange with neighbors.
        \item Metric evaluation and reputation updates.
        \item Weighted aggregation based on reputation.
    \end{itemize}
\end{itemize}

Each node maintains its own history of metrics and reputations, enabling smoothing, recurrence detection, and adaptive behavior. The integrated visualization tools help track the evolution of each metric, monitor node reputation across rounds, and observe the influence of distributed feedback.

\begin{figure}[htbp]
\centering
\includegraphics[width=0.4\textwidth]{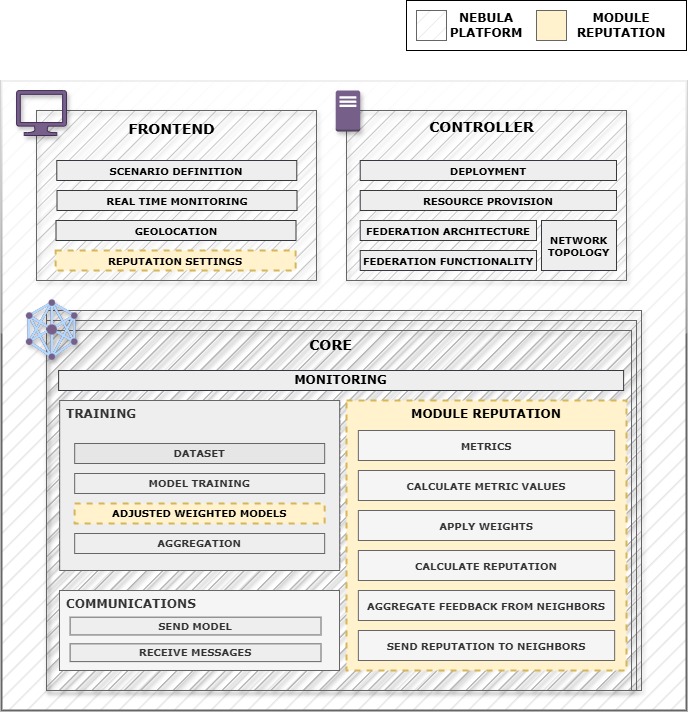}
\caption{Integration of the RepuNet reputation system into the Nebula platform architecture.}
\label{fig:nebula-final}
\end{figure}

\section{RepuNet Evaluation}
\label{sec:validation}

This section evaluates RepuNet in adversarial DFL environments. The objective is to assess its effectiveness in detecting and penalizing malicious nodes, preserving model performance and training stability. The experiments vary the proportion of attackers and the timing of activation, analyzing the evolution of reputation scores and their effect on aggregation, with and without defense.

\subsection{Experimental Environment}

Experiments were run on Nebula, which emulates DFL via virtual scenarios. Nodes train, evaluate, and aggregate locally without a central server. The entire federation runs as independent processes on a single machine, enabling full control and real-time monitoring.
Table~\ref{tab:experiment_parameters} summarizes the general setup parameters, including datasets, topologies, hardware configuration, and training schedule used across all experiments.

\begin{table}[htbp]
\centering
\caption{Summary of Experimental Setup Parameters}
\scriptsize
\renewcommand{\arraystretch}{1.2}
\label{tab:experiment_parameters}
\resizebox{\linewidth}{!}{%
\begin{tabular}{
    >{\raggedright\arraybackslash}m{2cm}
    >{\raggedright\arraybackslash}m{7cm}
}
\hline
\textbf{Parameter} & \textbf{Description} \\
\hline
Platform & Nebula (Decentralized FL emulation platform) \\
Datasets & MNIST, CIFAR-10 \\
Nodes per federation & 10--25 nodes \\
Topology types & Fully connected, Random \\
Data partition & Dirichlet (Non-IID), $\alpha \in \{0.1, 0.3, 0.5, 0.9\}$ \\
Local epochs & MNIST: 1 epoch/round; CIFAR-10: 10 epochs/round \\
Aggregation algorithm & FedAvg \\
Hardware specs & Intel Xeon CPU E5-2697, 62 GB RAM, Ubuntu 22.04 \\
Number of rounds & 20 communication rounds \\
Timeout for aggregation & 30--60 seconds \\
Metrics used & Model Similarity, Fraction of Parameters Changed, Arrival Latency, Incoming Message Flow \\
\hline
\end{tabular}
}
\end{table}

We emulated model poisoning, delayer, and flooding attacks to test robustness. For each, the experimental task and measurement objective are clearly defined in Table~\ref{tab:attack_objectives}.

\begin{table}[htbp]
\centering
\caption{Attack Types, Tasks, and Measurement Objectives}
\scriptsize
\renewcommand{\arraystretch}{1.2}
\label{tab:attack_objectives}
\resizebox{\linewidth}{!}{%
\begin{tabular}{
    >{\raggedright\arraybackslash}m{1.8cm}
    >{\raggedright\arraybackslash}m{2.1cm}
    >{\raggedright\arraybackslash}m{5cm}
}
\hline
\textbf{Attack Type} & \textbf{Task} & \textbf{Measurement Objective} \\
\hline
Model Poisoning & Image classification (MNIST, CIFAR-10) & Evaluate degradation of global model accuracy (F1-score) due to manipulated updates. \\
Delayer Attack & Image classification (MNIST) & Assess synchronization issues and increased aggregation latency caused by delayed submissions. \\
Flooding Attack & Image classification (MNIST) & Measure communication efficiency degradation and CPU overhead caused by excessive messaging. \\
\hline
\end{tabular}
}
\end{table}

\subsection{Model Poisoning Attack}

This attack degrades the global model by injecting malicious updates that differ significantly from honest contributions and may hinder convergence if undetected. RepuNet evaluates each neighbor before aggregation based on behavioral metrics.
Experiments were conducted using MNIST and CIFAR-10 datasets under different topologies, attacker proportions, and data heterogeneity levels. Table~\ref{tab:model_poisoning_scenarios} summarizes the configuration of all poisoning scenarios, including early and intermittent variants. Each scenario is labeled with a short identifier (e.g., \textit{Base}, \textit{2.1}, \textit{6.1}) used consistently across results and discussion.

\begin{table}[htbp]
\centering
\caption{Model poisoning attack scenarios: configuration summary.}
\renewcommand{\arraystretch}{1.2}
\scriptsize
\resizebox{\columnwidth}{!}{%
\begin{tabular}{
    >{\raggedright\arraybackslash}m{2.5cm}
    >{\centering\arraybackslash}m{2cm}
    >{\centering\arraybackslash}m{2cm}
    >{\centering\arraybackslash}m{2cm}
}
\hline
\textbf{Scenario Group} & \textbf{Topology} & \textbf{Malicious Nodes} & \textbf{Dirichlet. $\alpha$} \\
\hline
Base / 1.x / 2.x         & Fully             & 30--60\%                 & 0.5 \\
3.1                     & Ring              & 30\%                     & 0.5 \\
3.2                     & Random            & 30\%                     & 0.5 \\
4.1 / 4.2               & Fully             & 30\%                     & 0.3 / 0.9 \\
5.1 / 5.2 / 5.3         & Fully             & 30\%                     & CIFAR10: 0.5 / 0.1 \\
6.1                    & Fully             & 30\%                     & 0.5 (early) \\
7.1 / 7.2               & Fully             & 30\%                     & 0.5 (interval 2/3) \\
\hline
\end{tabular}
}
\label{tab:model_poisoning_scenarios}
\end{table}

Table~\ref{tab:combined_poisoning_start} reports the F1-score for each scenario with and without RepuNet. Scores are measured at round 8 and 11, respectively. The difference $\Delta F_1$ quantifies the improvement, and the last column categorizes the observed impact.

\begin{table}[htbp]
\centering
\caption{F1-score comparison across model poisoning scenarios with and without RepuNet.}
\renewcommand{\arraystretch}{1.2}
\scriptsize
\resizebox{\columnwidth}{!}{%
\begin{tabular}{
    >{\centering\arraybackslash}m{1cm}  
    >{\centering\arraybackslash}m{1.5cm}  
    >{\centering\arraybackslash}m{1.8cm}  
    >{\centering\arraybackslash}m{1.5cm}  
    >{\centering\arraybackslash}m{1.5cm}  
}
\hline
\textbf{Scenario} & \textbf{F1 w/rep (r8)} & \textbf{F1 w/o rep (r11)} & $\boldsymbol{\Delta F_1}$ & \textbf{Impact} \\
\hline
Base  & 0.6800 & 0.3610 & 0.3190 & High \\
1.1   & 0.6340 & 0.4226 & 0.2114 & Medium \\
1.2   & 0.6670 & 0.5868 & 0.0802 & Low \\
1.3   & 0.6447 & 0.5430 & 0.1017 & Medium \\
2.1   & 0.5972 & 0.4444 & 0.1528 & Medium \\
2.2   & 0.5051 & 0.3506 & 0.1545 & Medium \\
2.3   & 0.4143 & 0.1575 & 0.2568 & High \\
3.1   & 0.6821 & 0.4017 & 0.2804 & High \\
3.2   & 0.6744 & 0.3231 & 0.3513 & High \\
4.1   & 0.6736 & 0.5012 & 0.1724 & Medium \\
4.2   & 0.6787 & 0.5763 & 0.1024 & Medium \\
5.1   & 0.5360 & 0.3965 & 0.1395 & Medium \\
5.2   & 0.5122 & 0.4301 & 0.0821 & Low \\
5.3   & 0.5166 & 0.0720 & 0.4446 & High \\
6.1   & 0.6191 & 0.2225 & 0.3966 & High \\
7.1   & 0.6796 & 0.6833 & -0.0037 & Low \\
7.2   & 0.6879 & 0.6825 & 0.0054 & Low \\
\hline
\end{tabular}
}
\label{tab:combined_poisoning_start}
\end{table}

\begin{figure*}[htbp]
    \centering
    \subfloat[\scriptsize General scenarios\label{fig:impact_general}]{
    \includegraphics[width=0.28\linewidth]{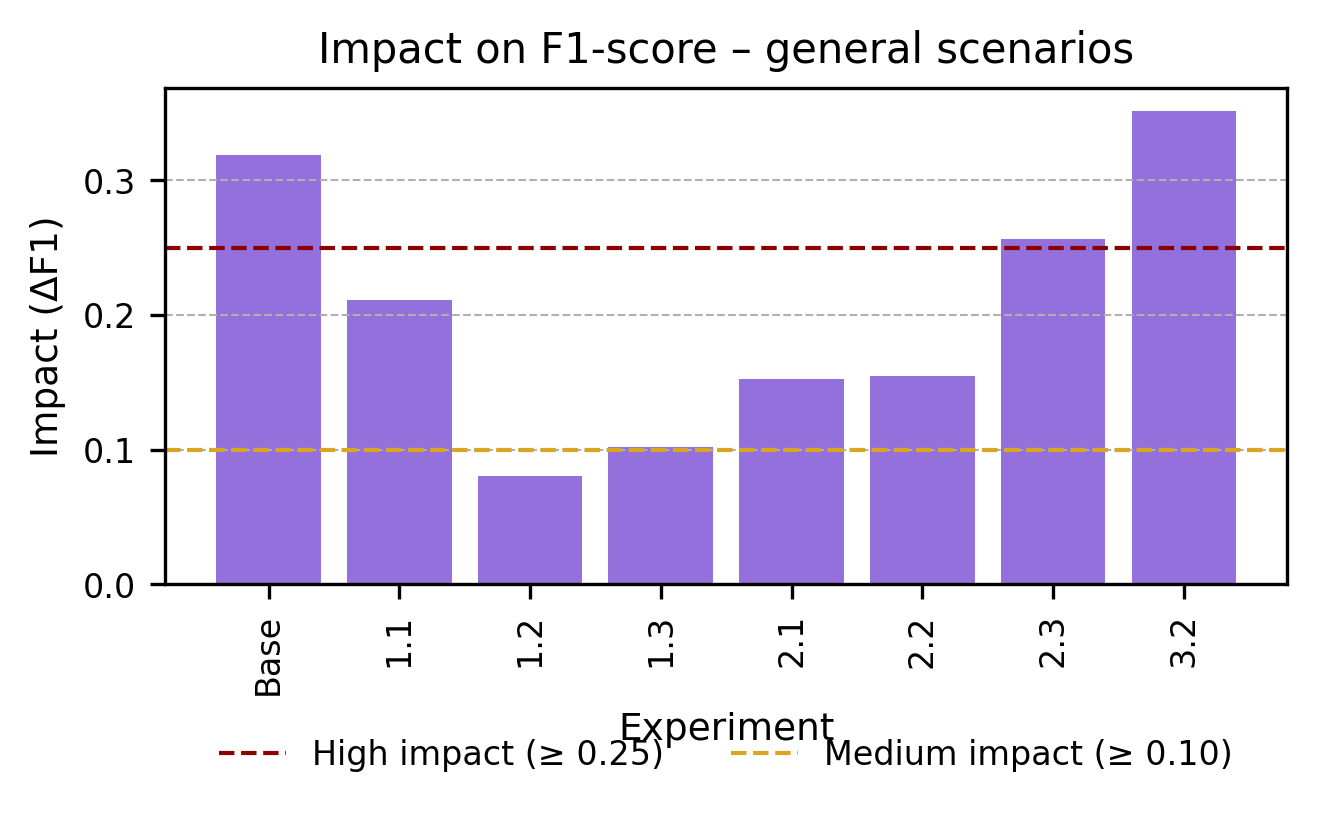}
    }
    \subfloat[\scriptsize Early attacks (round 1)\label{fig:impact_early}]{
        \includegraphics[width=0.28\linewidth]{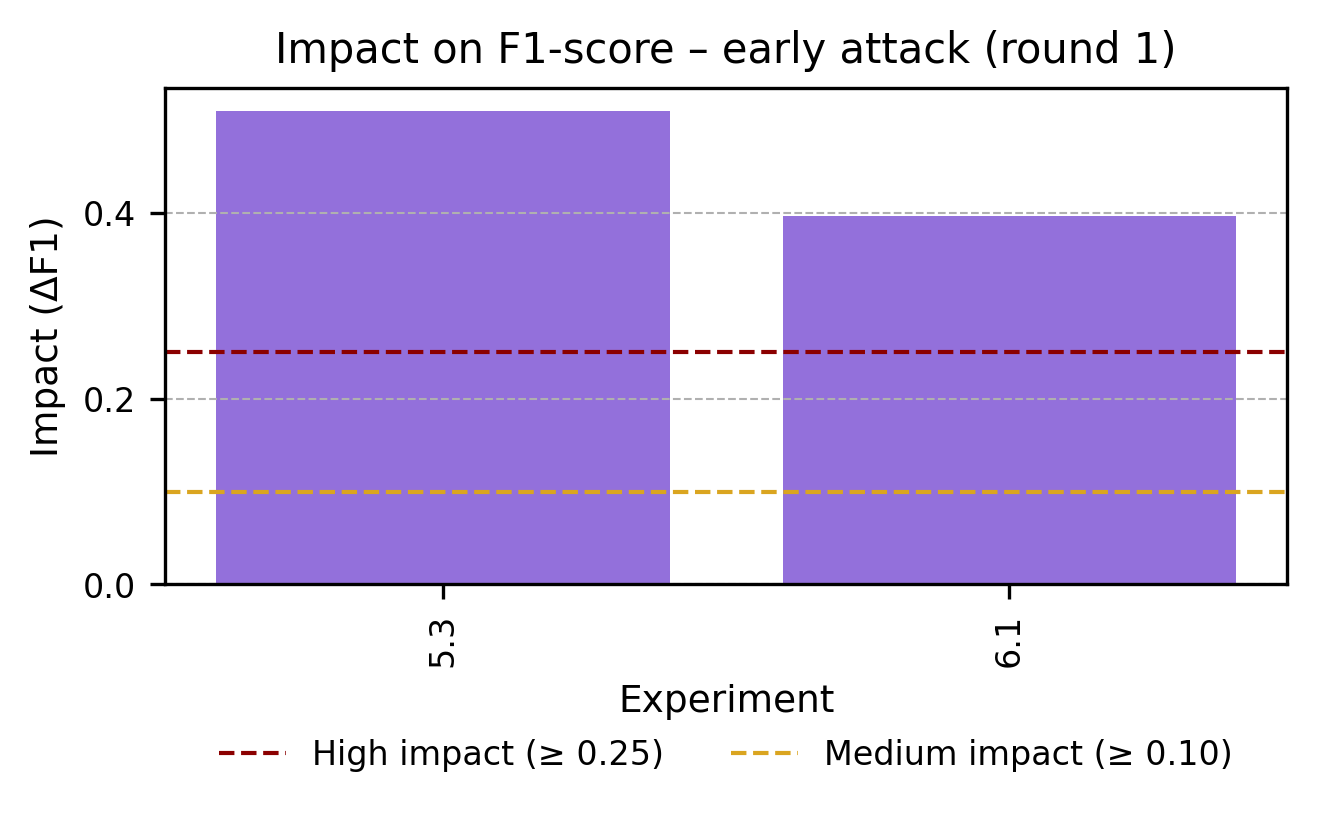}
    }
    \subfloat[\scriptsize Intermittent attacks\label{fig:impact_interval}]{
        \includegraphics[width=0.28\linewidth]{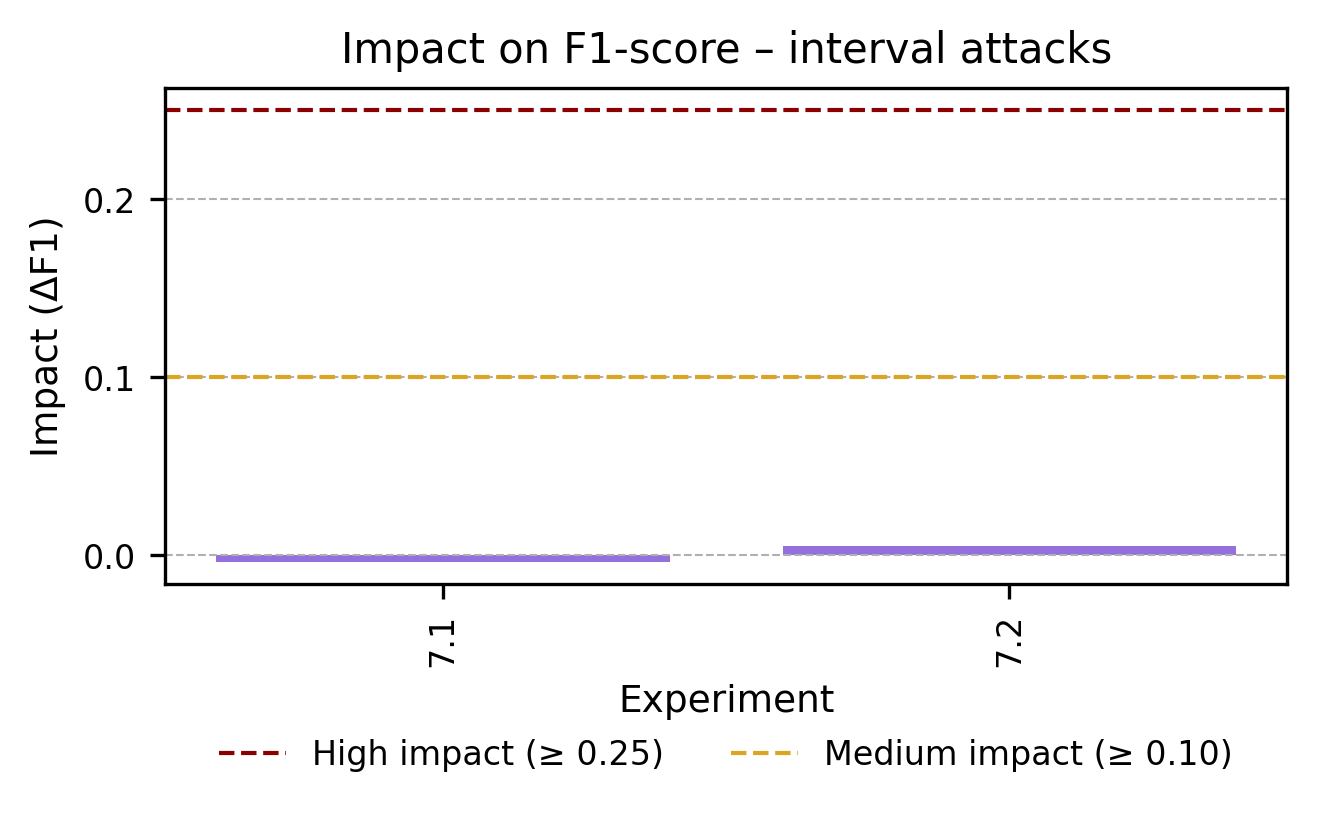}
    }
    \vspace{1em}
    \subfloat[\scriptsize Dirichlet $\alpha$ variation (4.1 vs 4.2)\label{fig:impact_dirichlet_1}]{
        \includegraphics[width=0.28\linewidth]{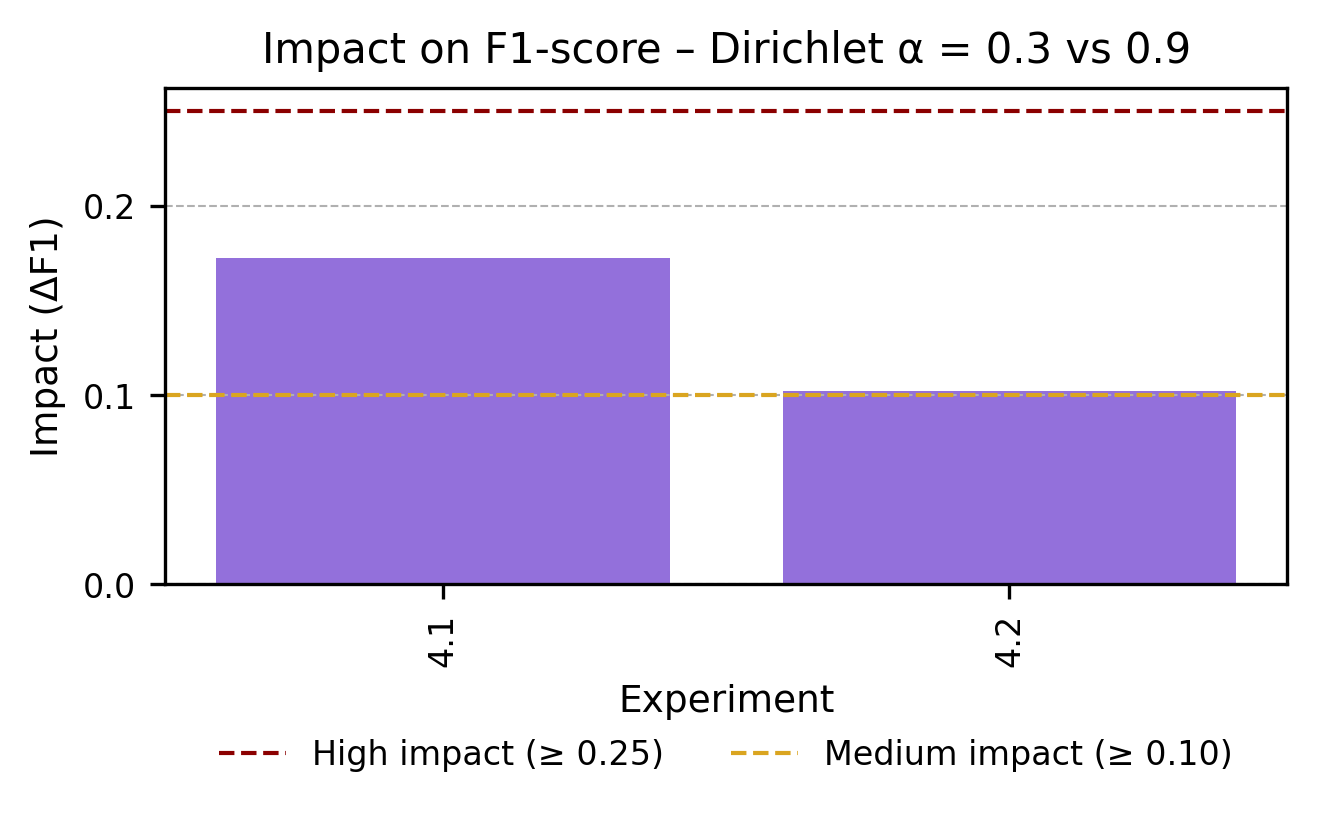}
    }
    \subfloat[\scriptsize Dirichlet $\alpha$ variation (5.1 vs 5.2)\label{fig:impact_dirichlet_2}]{
        \includegraphics[width=0.28\linewidth]{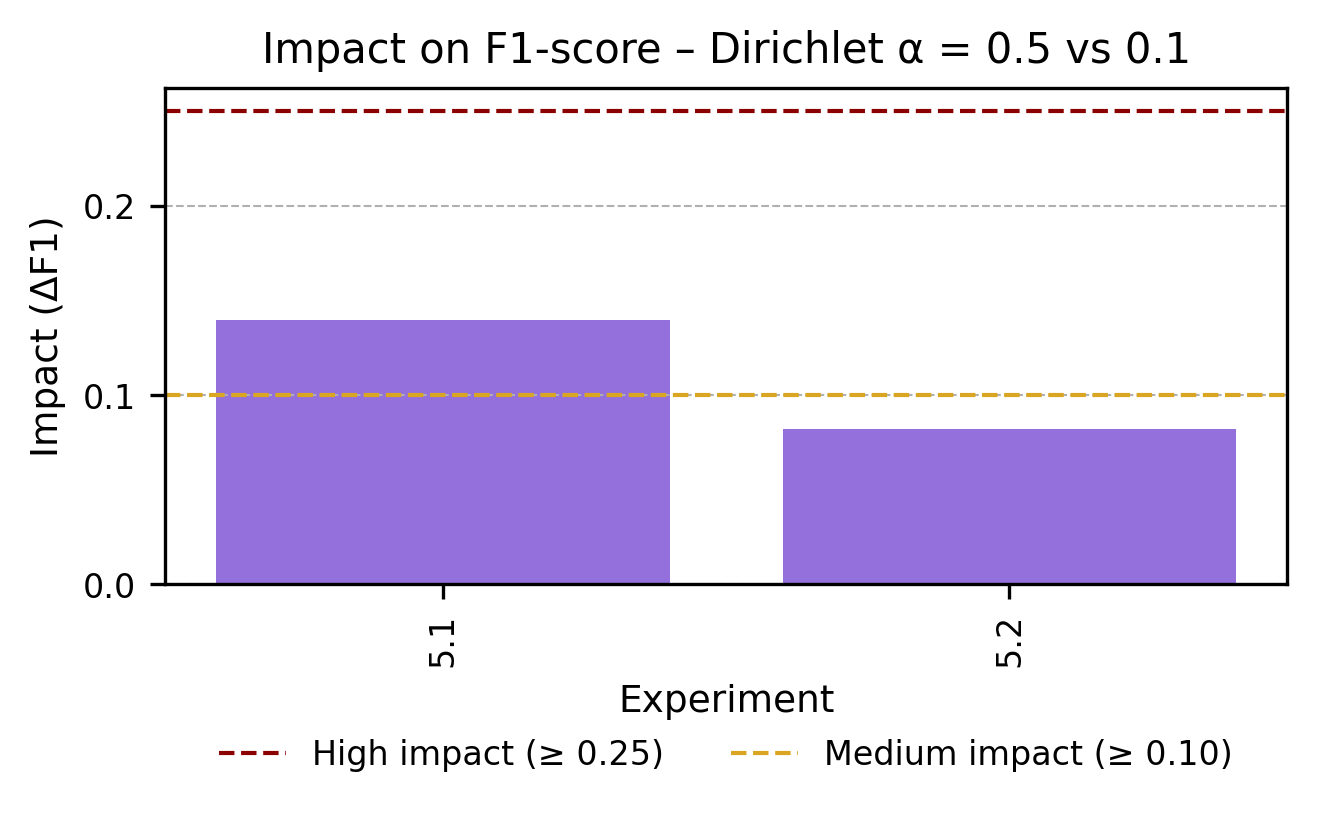}
    }
    \caption{Impact of the RepuNet reputation system in various \textit{model poisoning} attack scenarios. (a) General scenarios, (b) attacks from round 1, (c) intermittent attacks, (d--e) sensitivity to the heterogeneity parameter $\alpha$ of the Dirichlet distribution. Horizontal lines mark the thresholds for high impact ($\geq 0.25$) and medium impact ($\geq 0.10$).}
    \label{fig:impact_thresholds_summary}
\end{figure*}

\begin{figure}[htbp]
    \centering
    \captionsetup{font=small}

    \subfloat[Base – All nodes]{
        \includegraphics[width=0.32\linewidth]{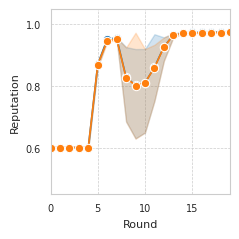}
    }
    \subfloat[3.2 – All nodes]{
        \includegraphics[width=0.32\linewidth]{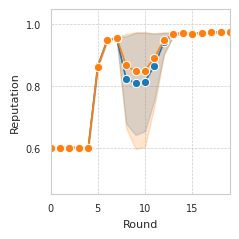}
    }
    \subfloat[2.3 – All nodes]{
        \includegraphics[width=0.32\linewidth]{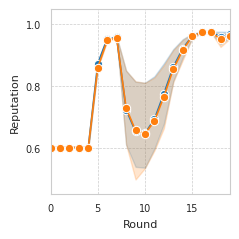}
    }
    \vspace{0.5mm}
    \subfloat[Base – Malicious nodes]{
        \includegraphics[width=0.32\linewidth]{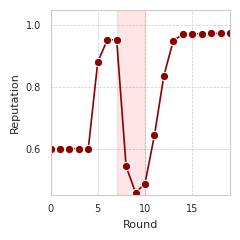}
    }
    \subfloat[3.2 – Malicious nodes]{
        \includegraphics[width=0.32\linewidth]{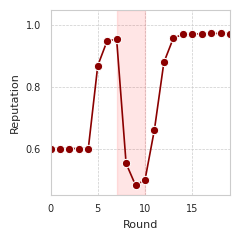}
    }
    \subfloat[2.3 – Malicious nodes]{
        \includegraphics[width=0.32\linewidth]{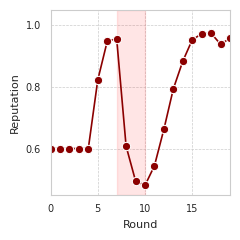}
    }

    \caption{Reputation evolution in three scenarios with \textit{model poisoning} attacks. Top row: average reputation of all federation nodes. Bottom row: average reputation of malicious nodes.}
    \label{fig:reputation_combined}
\end{figure}

Figure~\ref{fig:impact_thresholds_summary} provides a visual summary of the impact levels across all scenario groups. Subfigures organize experiments by activation type or configuration and highlight thresholds for low, medium, and high impact levels.
Three representative scenarios illustrate RepuNet’s response to poisoning attacks. These experiments use the MNIST dataset and simulate attacks in federations with 10 nodes. Each case includes two plots: one showing the average reputation of all nodes, and another focused on malicious nodes only. The results are shown in Figure~\ref{fig:reputation_combined}, which illustrates how reputation evolves in response to adversarial behavior under different topologies and attacker proportions.

Scenario \textbf{Base} represents a fully connected topology with 30\% malicious nodes. RepuNet responds quickly after the attack is triggered in round 7, penalizing adversarial behavior. The reputation of honest nodes remains high, while that of malicious nodes drops noticeably. Nodes 192.168.51.10:45009 and 192.168.51.11:45010 were selected for illustration.
Scenario \textbf{3.2} uses a random topology and maintains the same proportion of attackers. Greater variability in reputation is observed due to the network’s sparse connectivity, which reduces the influence of malicious nodes. The drop in reputation begins in round 7, confirming RepuNet’s ability to adapt in less structured environments. Nodes 192.168.51.2:45001 and 192.168.51.9:45008 are shown.
Scenario \textbf{2.3} corresponds to a more adversarial environment, with a fully connected topology and 60\% malicious nodes. Despite the increased pressure, RepuNet still penalizes malicious behavior consistently. The Dirichlet parameter $\alpha=0.5$ used for data partitioning has limited effect in this setting. Nodes 192.168.51.4:45003 and 192.168.51.9:45008 were selected for observation.
To illustrate the internal adjustment process, Table~\ref{tab:nodo10_6_short} shows how node 10 tracks a malicious neighbor across rounds. In rounds 7 and 8, after the attack starts, the model similarity and fraction of parameters changed drop significantly. This leads to an automatic reweighting, increasing the importance of these more discriminative metrics. The system adapts to contextual anomalies, penalizing malicious behavior without manual intervention.

\begin{table}[htbp]
\centering
\caption{Metric and weight evolution for a malicious neighbor (node 10).}
\renewcommand{\arraystretch}{1.5}
\resizebox{\columnwidth}{!}{%
\begin{tabular}{
    >{\raggedright\arraybackslash}m{2.5cm}
    >{\centering\arraybackslash}m{1.2cm}
    >{\centering\arraybackslash}m{1.2cm}
    >{\centering\arraybackslash}m{1.2cm}
    >{\centering\arraybackslash}m{1.2cm}
}
\hline
\textbf{Metric} & \textbf{Round 5} & \textbf{Round 6} & \textbf{Round 7} & \textbf{Round 8} \\
\hline
Model similarity & 0.84 & 0.82 & 0.42 & 0.36 \\
Fraction of parameters changed & 0.79 & 0.75 & 0.33 & 0.29 \\
Arrival latency & 0.91 & 0.93 & 0.89 & 0.94 \\
Messages exchanged & 0.86 & 0.88 & 0.85 & 0.87 \\
\hline
\textbf{Weight: similarity} & 0.25 & 0.24 & 0.41 & 0.43 \\
\textbf{Weight: fraction} & 0.22 & 0.22 & 0.38 & 0.39 \\
\textbf{Weight: latency} & 0.28 & 0.28 & 0.11 & 0.10 \\
\textbf{Weight: messages} & 0.25 & 0.26 & 0.10 & 0.08 \\
\hline
\end{tabular}
}
\label{tab:nodo10_6_short}
\end{table}

\subsection{Delayer Attack}

The \textit{delayer} attack disrupts the federated process by intentionally delaying the transmission of model updates. While it does not alter model content, it exploits the temporal dimension—latency—to desynchronize nodes, degrade convergence, and stall training.
Experiments on MNIST were conducted using the Nebula platform, exploring different delay durations, topologies, and attacker proportions. \textbf{Table~\ref{tab:delay_scenarios}} summarizes the configurations evaluated.

\begin{table}[htbp]
\centering
\caption{Summary of delay attack configurations tested.}
\renewcommand{\arraystretch}{1.2}
\scriptsize
\begin{tabular}{
    >{\raggedright\arraybackslash}m{2cm}
    >{\centering\arraybackslash}m{1cm}
    >{\centering\arraybackslash}m{1cm}
    >{\centering\arraybackslash}m{2cm}
    >{\centering\arraybackslash}m{1cm}
}
\hline
\textbf{Scenario Group} & \textbf{Topology} & \textbf{Delay (s)} & \textbf{Malicious Nodes} & \textbf{Interval} \\
\hline
Base / 2.x              & Fully   & 20            & 30--60\%          & 1 \\
3.2                     & Random  & 30            & 30\%              & 1 \\
4.x                     & Fully   & 40--60        & 30\%              & 1 \\
5.x (intermittent)      & Fully   & 80            & 30\%              & 2--4 \\
6.x                     & Fully   & 100           & 30--60\%          & 1 \\
\hline
\end{tabular}
\label{tab:delay_scenarios}
\end{table}

\begin{figure}[htbp]
    \centering
    \captionsetup{font=small}

    \subfloat[Global reputation – attack from round 7\label{fig:avg_rep_r7}]{
        \includegraphics[width=0.45\linewidth]{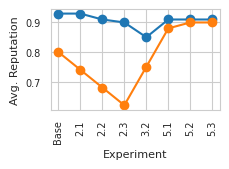}
    }
    \hfill
    \subfloat[Malicious reputation – attack from round 7\label{fig:avg_rep_r7_mal}]{
        \includegraphics[width=0.45\linewidth]{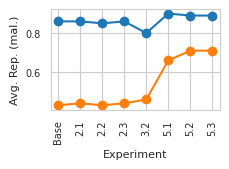}
    }

    \vspace{1.5mm}

    \subfloat[Global reputation – attack from round 1\label{fig:avg_rep_r1}]{
        \includegraphics[width=0.45\linewidth]{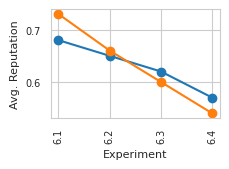}
    }
    \hfill
    \subfloat[Malicious reputation – attack from round 1\label{fig:avg_rep_r1_mal}]{
        \includegraphics[width=0.45\linewidth]{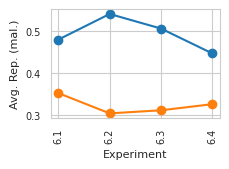}
    }

    \vspace{1.5mm}

    \subfloat[Global reputation – intermittent attack\label{fig:avg_rep_r7_interval}]{
        \includegraphics[width=0.45\linewidth]{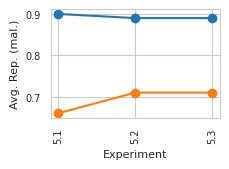}
    }
    \hfill
    \subfloat[Malicious reputation – intermittent attack\label{fig:avg_rep_r7_mal_interval}]{
        \includegraphics[width=0.45\linewidth]{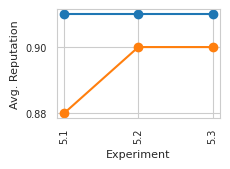}
    }

    \caption{Evolution of average reputation across different \textit{delayer} attack configurations. Each row compares global reputation (left) and malicious node reputation (right) under a specific attack timing: from round 7, from round 1, and with intermittent activation.}
    \label{fig:avg_reputation_summary}
\end{figure}

\begin{figure*}[htbp]
    \centering
    \includegraphics[width=1\linewidth]{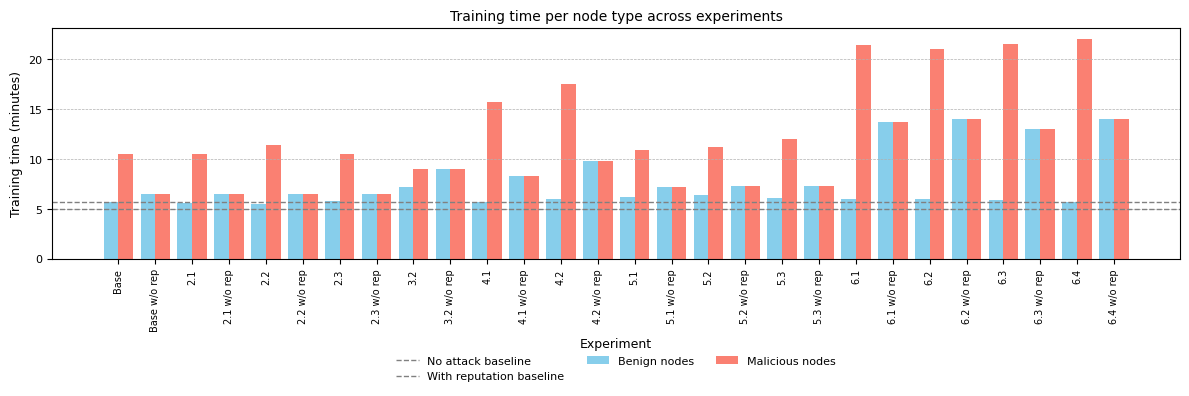}
    \caption{Total training time per node type in the different scenarios.}
    \label{fig:total_time_grouped}
\end{figure*}

Unlike poisoning, this attack targets efficiency. Figure~\ref{fig:total_time_grouped} presents the total training time by node type, showing how lag propagates and increases desynchronization when no mitigation is applied. For reference, the baseline training duration without attack was 5 minutes; with reputation enabled, it increased slightly to 5.7 minutes.
To assess how reputation adapts, Figure~\ref{fig:avg_reputation_summary} shows average reputation evolution under different attack timings. Figures (a) and (b) correspond to the global reputation for attacks starting in round 7 and round 1, respectively. Figures (c) and (d) show the reputation of malicious nodes in the same scenarios. Finally, (e) and (f) present global and malicious reputations under intermittent attacks with intervals 2, 3, and 4.

\begin{figure}[htbp]
    \centering
    \captionsetup{font=small}

    \subfloat[Round 2 – attack from round 1]{
        \includegraphics[width=0.42\linewidth]{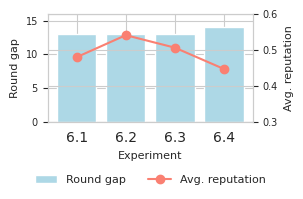}
    }
    \hfill
    \subfloat[Round 9 – attack from round 7]{
        \includegraphics[width=0.48\linewidth]{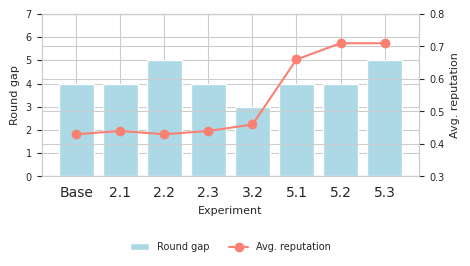}
    }

    \caption{Relationship between the round gap and average reputation of malicious nodes in two scenarios.}
    \label{fig:gap_vs_rep}
\end{figure}

Figure~\ref{fig:gap_vs_rep} shows round gaps between benign and malicious nodes. Subfigure (a) refers to scenarios where the attack begins in round 1; (b) shows scenarios with attack starting from round 7. In both cases, average reputation of malicious nodes is included for correlation.

\begin{figure}[htbp]
    \centering
    \captionsetup{font=small}

    \subfloat[Round 9 – attack from round 7]{
        \includegraphics[width=0.42\linewidth]{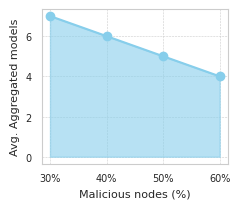}
    }
    \hfill
    \subfloat[Round 2 – attack from round 1]{
        \includegraphics[width=0.42\linewidth]{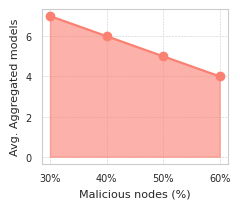}
    }

    \caption{Average number of models aggregated by percentage of malicious nodes.}
    \label{fig:agg_models}
\end{figure}

Lastly, Figure~\ref{fig:agg_models} reports the average number of aggregated models depending on the percentage of malicious nodes. Panel (a) refers to round 9 (attack from round 7), while (b) shows round 2 (attack from round 1). These results demonstrate how model aggregation is reduced in the presence of delays.

\subsection{Flooding Attack}

The \textit{flooding} attack aims to saturate the communication layer by overwhelming the network with a high volume of messages from malicious nodes. This behavior degrades communication efficiency, increases CPU overhead, and can ultimately impact model convergence. RepuNet addresses this threat by penalizing nodes that exhibit abnormal message volume.
Table~\ref{tab:flooding_scenarios} summarizes the configurations evaluated for this attack, including variations in topology, proportion of malicious nodes, activation round, and message injection interval.

\begin{table}[htbp]
\centering
\caption{Flooding attack scenarios: configuration summary (grouped).}
\renewcommand{\arraystretch}{1.2}
\scriptsize
\resizebox{\columnwidth}{!}{%
\begin{tabular}{
    >{\raggedright\arraybackslash}m{2.2cm}
    >{\centering\arraybackslash}m{2.2cm}
    >{\centering\arraybackslash}m{2.2cm}
    >{\centering\arraybackslash}m{1.5cm}
}
\hline
\textbf{Scenario Group} & \textbf{Topology} & \textbf{Malicious Nodes} & \textbf{Start Round} \\
\hline
Base / 2.x              & Fully             & 30--60\%                 & 7 \\
3.1                    & Random            & 30\%                     & 7 \\
4.1                    & Fully             & 30\%                     & 7 (interval 2) \\
5.x                    & Fully             & 30--60\%                 & 1 \\
\hline
\end{tabular}
}
\label{tab:flooding_scenarios}
\end{table}

\begin{figure}[htbp]
    \centering
    \captionsetup{font=small}
    \includegraphics[width=0.7\linewidth]{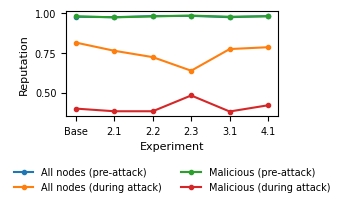}
    \caption{Average reputation evolution of all and malicious nodes during the flooding attack (activated in round 7).}
    \label{fig:flooding_reputation_evolution}
\end{figure}

Figure~\ref{fig:flooding_reputation_evolution} shows the evolution of the average reputation under flooding scenarios starting at round 7. Malicious nodes exhibit a sharp and consistent decline in reputation, while benign participants maintain high values, highlighting RepuNet’s ability to penalize anomalies without affecting honest contributions.

Additional scenarios were tested where the flooding attack was activated from the first round and ran until round 10. In these early-activation cases, round 0 served as the only observation phase, with all nodes initialized to a default reputation of 0.6. Table~\ref{tab:flooding_r1} presents the reputation scores observed in these scenarios.

\begin{table}[htbp]
\centering
\caption{Benign and malicious reputation under \textit{flooding} attacks starting at round 1.}
\renewcommand{\arraystretch}{1.2}
\scriptsize
\resizebox{\columnwidth}{!}{%
\begin{tabular}{
    >{\centering\arraybackslash}m{1cm}
    >{\centering\arraybackslash}m{2cm}
    >{\centering\arraybackslash}m{2cm}
    >{\centering\arraybackslash}m{2.2cm}
    >{\centering\arraybackslash}m{2.2cm}
}
\hline
\textbf{Scenario} & \textbf{Benign Rep. (r2)} & \textbf{Benign Rep. (r6)} & \textbf{Malicious Rep. (r2)} & \textbf{Malicious Rep. (r6)} \\
\hline
5.1 & 0.9387 & 0.9568 & 0.3256 & 0.2718 \\
5.2 & 0.9385 & 0.9347 & 0.2936 & 0.2970 \\
5.3 & 0.9050 & 0.9335 & 0.3025 & 0.3239 \\
5.4 & 0.9168 & 0.9313 & 0.2950 & 0.3914 \\
\hline
\end{tabular}
}
\label{tab:flooding_r1}
\end{table}

\begin{figure}[htbp]
    \centering
    \captionsetup{font=small}
    \subfloat[Attack from round 7]{
        \includegraphics[width=0.45\linewidth]{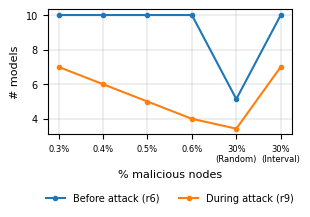}
    }
    \hfill
    \subfloat[Attack from round 1]{
        \includegraphics[width=0.45\linewidth]{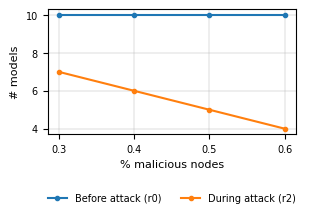}
    }
    \caption{Average number of models aggregated in round 9 under different flooding attack activation points. The Y-axis represents the average number of models aggregated by the federation nodes. The system consistently reduces the inclusion of malicious contributions as their proportion increases.}
    \label{fig:modelos_agregados}
\end{figure}

The impact on the aggregation process was also assessed by measuring the average number of models accepted in round 9. As shown in Figure~\ref{fig:modelos_agregados}, the number of aggregated models decreases proportionally with the increase in malicious nodes. This behavior holds for both early and delayed attack activation, confirming RepuNet’s robustness even with limited prior information.

Finally, to evaluate the indirect cost of flooding attacks, Table~\ref{tab:cpu_variacion_flooding} reports the variation in average CPU usage observed in benign nodes. This variation reflects the computational overhead caused by processing excessive messages. Scenarios with higher percentages of malicious nodes tend to exhibit increased CPU load, underscoring the importance of mitigating such attacks not only to preserve aggregation quality but also to reduce resource consumption.

\begin{table}[htbp]
\centering
\caption{CPU usage variation in benign nodes under \textit{flooding} attacks.}
\renewcommand{\arraystretch}{1.2}
\scriptsize
\resizebox{\linewidth}{!}{%
\begin{tabular}{
    >{\centering\arraybackslash}m{2.3cm}
    >{\centering\arraybackslash}m{2.3cm}
    >{\centering\arraybackslash}m{2.7cm}
}
\hline
\textbf{Scenario} & \textbf{Attack round} & \textbf{Avg. CPU Variation (\%)} \\
\hline
Base   & 7--12 & 5.23 \\
2.1    & 7--12 & -1.03 \\
2.2    & 7--12 & 15.14 \\
2.3    & 7--12 & 3.70 \\
3.1    & 7--12 & 11.21 \\
4.1    & 7--12 & -4.49 \\
5.1    & 1--10 & -8.14 \\
5.2    & 1--10 & 5.70 \\
5.3    & 1--10 & -4.80 \\
5.4    & 1--10 & -9.90 \\
\hline
\end{tabular}
}
\label{tab:cpu_variacion_flooding}
\end{table}

\section{Discussion}
\label{sec:discussion}

The experimental evaluation demonstrates that RepuNet provides effective protection against diverse adversarial behaviors in DFL scenarios. Rather than reiterating the internal mechanics of the reputation engine, this section focuses on analyzing its observed behavior, highlighting strengths, limitations, and areas for improvement.
Across all experiments, RepuNet consistently reduced the influence of malicious nodes during aggregation without compromising the participation structure of the network. Reputation trajectories showed rapid penalization after attack activation, confirming the system’s responsiveness. However, certain patterns—such as intermittent or oscillating attacks—exposed scenarios where nodes could recover influence too quickly, suggesting that stricter memory-based penalization strategies may be needed.
The ability to adapt dynamically to contextual changes, without centralized control or static heuristics, positioned RepuNet as a flexible defense strategy. Nonetheless, further refinement of metric weighting and exclusion thresholds could enhance robustness under more stealthy or coordinated threats. The following subsections delve into specific attack types and their mitigation outcomes.

\subsection{Model Poisoning}

Table~\ref{tab:combined_poisoning_start} presents the F1-scores obtained with and without RepuNet across several poisoning scenarios. The most significant improvements were observed in scenarios \textit{Base}, \textit{3.2}, and \textit{6.1}, where RepuNet increased performance by over 30 points in F1-score, reaching differences above 0.35.
In early-activation cases such as \textit{5.3}, where attacks began at round 1, the gap was even greater: F1-score improved from 0.0720 to 0.5166. This demonstrates RepuNet’s capacity to recover learning even under high pressure and without a prior observation phase.
Scenarios with skewed data distributions (e.g., \textit{5.2} with $\alpha = 0.1$) showed lower gains, as expected under higher heterogeneity, but still maintained improvements over the baseline. In intermittent scenarios (\textit{7.1}, \textit{7.2}), RepuNet showed limited effect, suggesting the system could benefit from memory-based metrics to counter low-frequency adversarial patterns.
Overall, the results confirm that RepuNet substantially mitigates poisoning attacks across various conditions, adapting both to topology and distribution challenges.

\subsection{Delayer Attack}

As shown in Table~\ref{tab:delay_scenarios}, the delayer attack was tested under various delays (20–100 seconds), attacker proportions, and topologies. RepuNet successfully penalized delayed nodes across all configurations. In early rounds (r2), scenarios such as \textit{2.3} and \textit{6.1} already showed reputation gaps of more than 0.3 between benign and malicious participants.
Reputation remained low for attackers in both fully connected and random topologies, as shown in Figure~\ref{fig:avg_reputation_summary}, confirming the system’s capacity to suppress desynchronizing behavior.
Intermittent delay scenarios revealed partial reintegration between inactive phases, which may be addressed by increasing the exclusion threshold or applying temporal penalties. Despite this, overall training time remained stable for honest nodes, while malicious participants completed fewer rounds (Figure~\ref{fig:total_time_grouped}).
The reduction in model aggregation under delay conditions was also consistent. Figure~\ref{fig:agg_models} shows that as the delay severity or attacker proportion increased, the number of models selected decreased accordingly—particularly in scenarios \textit{4.1} and \textit{6.3}.

\subsection{Flooding Attack}

Table~\ref{tab:flooding_scenarios} summarizes the configurations evaluated, including topology, proportion of malicious nodes, activation round, and interval. When the flooding attack was activated at round 7 (scenarios \textit{Base} to \textit{4.1}), RepuNet responded effectively: the reputation of malicious nodes dropped from initial values near 0.98 to values between 0.38 and 0.48, while honest nodes maintained values around 0.80.
In early-activation scenarios (round 1, \textit{5.1}--\textit{5.4}), all nodes started from a neutral reputation of 0.6. Table~\ref{tab:flooding_r1} shows the evolution of reputation for benign and malicious nodes. Benign nodes rapidly recovered to values above 0.93 by round 6, while adversarial nodes remained low or exhibited only limited recovery.
The number of aggregated models in round 9 (Figure~\ref{fig:modelos_agregados}) decreased progressively with the proportion of malicious nodes, both for early and delayed attacks, indicating consistent exclusion of unreliable contributions.
Intermittent attacks such as \textit{4.1} revealed partial reintegration of malicious nodes between flooding bursts, suggesting a limitation of short-term metrics. Despite this, RepuNet re-penalized such behavior in subsequent rounds, preserving overall robustness.
Table~\ref{tab:cpu_variacion_flooding} presents CPU usage variations for benign nodes. In most cases, RepuNet reduced or stabilized overhead, with the highest efficiency gains observed in scenarios \textit{5.1} and \textit{5.3}. These results confirm that reputation-based filtering improves not only model quality but also resource efficiency under flooding conditions.

\section{Conclusions}
\label{sec:conclusion}

This work presented a reputation mechanism for DFL that detects and mitigates malicious behavior during training. The system evaluates each node based on locally observable metrics—model similarity, parameter changes, latency, and message volume—and updates reputation scores dynamically. Nodes with low reputation are excluded from aggregation, while recovery is possible upon improved behavior. Experiments confirm the system’s effectiveness against model poisoning, delay, and flooding attacks. In all cases, malicious nodes suffered a sharp drop in reputation after initiating the attack, enabling their exclusion without degrading the aggregated model.

RepuNet could be extended to deal with threats like data poisoning, which opens possibilities for integrating new metrics that assess model coherence with local data. Future improvements could also include a dynamic exclusion threshold, instead of a fixed one, and improved weight assignment by analyzing Z-score normalization, Bayesian schemes, or attention mechanisms to adjust metric relevance based on context.

\section*{Acknowledgment}
This work was supported by \textit{(a)} the Swiss Federal Office for Defense Procurement (armasuisse) with DECIMAL project (CYD-C-2020003), \textit{(b)} 21629/FPI/21, Fundación Séneca, Región de Murcia (Spain), \textit{(c)} the strategic project DEFENDER from the Spanish National Institute of Cybersecurity (INCIBE), by the Recovery, Transformation and Resilience Plan, Next Generation EU, and \textit{(d)} the European Commission through the Horizon Europe/JU SNS project ROBUST-6G (Grant Agreement no. 101139068).

\bibliographystyle{IEEEtran}
\bibliography{references}
\begin{IEEEbiographynophoto}{Isaac Marroquí Penalva}
received his M.Sc. degree in Software Engineering from the University of Murcia, Spain, in 2023. His master's thesis focused on the design and evaluation of reputation mechanisms for decentralized federated learning systems, with applications in cybersecurity and the Internet of Things. His current research interests include trust management, collaborative learning, and secure distributed systems.
\end{IEEEbiographynophoto}
\vskip -2.7\baselineskip plus -1fil
\begin{IEEEbiographynophoto}{Enrique Tomás Martínez Beltrán} is working towards a Ph.D. in Computer Science at the University of Murcia, Spain. His research interests include cybersecurity, IoT, and collaborative learning using FL.
\end{IEEEbiographynophoto}
\vskip -2.7\baselineskip plus -1fil
\begin{IEEEbiographynophoto}{Manuel Gil Perez} is an Associate Professor in the Department of Information and Communication Engineering of the University of Murcia, Murcia, Spain. His scientific activity is mainly devoted to cyber security, including intrusion detection systems, trust management, privacy-preserving data sharing, and security operations in highly dynamic scenarios. 
\end{IEEEbiographynophoto}
\vskip -2.7\baselineskip plus -1fil
\begin{IEEEbiographynophoto}{Alberto Huertas Celdrán} is an assistant professor at the University of Murcia and a guest researcher at the Communication Systems Group CSG, Department of Informatics IfI, University of Zürich. He received his PhD in Computer Science from the University of Murcia, Spain. His scientific interests include cybersecurity, FL, and computer networks.
\end{IEEEbiographynophoto}
\vfill

\end{document}